\documentclass[useAMS,usenatbib]{mn2e}
\usepackage{graphicx}
\usepackage{txfonts}
\usepackage{natbib}
\newcommand{\xte}{{\it RXTE}}

\newcommand{\xmm}{{\it XMM-Newton}}

\newcommand{\chandra}{{\it Chandra}}
\newcommand{\sax}{{\it BeppoSAX}}

\newcommand{\asca}{{\it ASCA}}

\newcommand{\suzaku}{{\it Suzaku}}
\def\gsim{\mathrel{\hbox{\rlap{\hbox{\lower4pt\hbox{$\sim$}}}\hbox{$>$}}}}
\def\lsim{\mathrel{\hbox{\rlap{\hbox{\lower4pt\hbox{$\sim$}}}\hbox{$<$}}}}

\begin{document}
   \title[On the evidence for narrow, relativistically shifted X-ray lines]{On the evidence for narrow, relativistically shifted X-ray lines}

   \author[S. Vaughan and P. Uttley]{S. Vaughan$^{1}$\thanks{email: sav2@star.le.ac.uk} and P. Uttley$^{2}$\\
   $^{1}$ X-Ray and Observational Astronomy Group, University of
   Leicester,  Leicester, LE1 7RH, U.K.\\
   $^{2}$ School of Physics and Astronomy, University of Southampton,
   Southampton SO17 1BJ, UK
   }

   \date{Accepted 2008 July 28. Received 2008 June 17; in original form 2008 April 18}

   \pagerange{\pageref{firstpage}--\pageref{lastpage}} \pubyear{2002}

   \maketitle
   
   \label{firstpage}

   \begin{abstract}
   In recent years there have been many reported detections of highly redshifted or blueshifted narrow spectral lines (both emission or absorption) in the X-ray spectra of active galaxies, but these are all modest detections in terms of their statistical significance. The aim of this paper is to review the issue of the significance of these detections and, in particular, take account of publication bias. A literature search revealed $38$ reported detections of narrow, strongly shifted ($v/c \ge 0.05$) X-ray lines in the $1.5-20$~keV spectra of Seyfert galaxies and quasars. These published data show a close, linear relationship between the estimated line strength and its uncertainty, in the sense that better observations (with smaller uncertainties) only ever show the smallest lines. This result is consistent with many of the reported lines being false detections resulting from random fluctuations, drawn from a large body of data and filtered by publication bias such that only the most `significant' fluctuations are ever reported. The reality of many of these features, and certainly their prevalence in the population at large, therefore remains an open question that is best settled though uniform analysis (and reporting) of higher quality observations.
   \end{abstract}

   \begin{keywords}
     galaxies: active --
     methods: data analysis --
     methods: statistical --
     X-rays: general --
     X-rays: galaxies
     \end{keywords}
 

\section{Introduction}
\label{sect:intro}

The arrival in recent years of large quantities of X-ray data from CCD and grating spectrometers has provided a vast increase in the amount of spectral information available to X-ray astronomers. These data have vastly improved our understanding of the X-ray properties of active galaxies such as Seyfert galaxies and quasars. Among the discoveries to have been made using these data were narrow line-like features in emission or absorption at unexpected energies in the canonical $2-10$  keV X-ray band \citep[see e.g.][]{yaqoob99,turner02,pounds03a,porquet04,matt05,longi06,cappi06}. In many cases these have been identified with transitions in iron that have been strongly redshifted or blueshifted out of the usual iron line band ($6.4-6.9$ keV) by high bulk flow velocities ($v \gsim 0.05c$) or gravitational effects. Yet they are narrow and line-like, indicating low velocity dispersion. The narrowness and highly shifted centroid energies of these features mark them as distinct from the relativistically broadened emission lines seen in Seyfert galaxies \citep[see e.g.][for reviews]{fabian00,reynolds03}. The extreme nature of these features means they have profound implications for the structure and energetics of the nucleus \citep{cappi06}. 

Irrespective of their proposed physical importance the reported features are all rather modest detections, in the sense that the statistical significance is not outstanding in any single case. Indeed, the vast majority of reported cases appear to lie in the ``$2-3\sigma$'' regime\footnote{Usually this means the outcome of an hypothesis test was a $p$-value in the range from $\sim 5\times 10^{-2}$ down to $\sim 3 \times 10^{-3}$. Such results are usually reported as ``detected at $95 - 99.7$ per cent confidence'' or given in units of $\sigma$ by comparison with the tail area under the Normal curve.}. The detection process usually involves modelling the featureless continuum emission of the target object and searching for large, localised, positive or negative residuals between the data and model (spectral lines may appear in emission or absorption); the larger the contribution of the residuals to the fit statistic (e.g. the change in $\chi^2$ upon including a line in the fitted model) the more significant the feature. The wide bandpass and good spectral resolution of modern detectors, combined with the number of datasets that have been processed mean that data archives for recent spectroscopy missions will  contain many modest signal-to-noise ``features'' simply from random sampling fluctuations in the photon counting signal from otherwise featureless continua. For example, the Tartarus database\footnote{{\tt http://astro.ic.ac.uk/Research/Tartarus/}} contains some $661$ \asca\ observations of active galaxies, and the \xmm\ Science Archive (XSA\footnote{{\tt http://xmm.esac.esa.int/xsa/index.shtml}}) contains over 1,000 publically available ``Guest Observer'' observations listed under the proposal category ``AGNs, QSOs, BL Lacs and XRB''. The number of spectra is arguably even higher than this because each of the longer observations is often divided into multiple spectra as a function of time, source flux, and so on, and indeed many of the line detection papers report the features to be ``transient'' (i.e. detected in only a subset of the data). 

The large number of moderately significant detections (see sect.~\ref{sect:analysis}) might be considered as evidence to support the reality of the features. However, when considering results presented for individual datasets drawn from a much larger population of available data one should be aware of the distorting effects of ``publication bias,'' also known as the ``file-drawer effect'' -- the tendency for positive results to be published and negative results to go unreported (``filed away''). See \citet{sterling59}, \citet{rosenthal79} and \citet{begg88} for general discussion of publication bias, and also \citet{stern97} and \citet{naylor97} for a more recent discussion of the importance of publication bias in the context of medical trials. One way to test for the presence of publication bias is through a ``funnel plot,'' originally proposed to aid meta-analyses of medical trials \citep{egger97}, which compares the size of a trial\footnote{The size of the sample used in the study; larger trials, i.e. those with larger sample sizes, tend to give higher signal-to-noise results.} to its estimate of the strength of the effect.  All estimates of the strength of an effect should be symmetrically scattered around the true value, with smaller trials providing less precise estimates and so larger scatter. In the absence of publication bias this will result in a symmetric funnel-shaped plot because the  estimates of the effect strength are independent of the sample sizes, but the scatter is larger for smaller samples. If publication bias is present, experiments or observations are less likely to be published if the estimate of the effect strength is low (or of low significance), i.e. the bias is against publishing non-detections, leading to an asymmetric funnel plot in which the strength of the effect is correlated with the sample size. The funnel plot produced from biased literature is the same as from the equivalent unbiased literature but with less ``interesting'' results (i.e. less significant or less strong results, which lie on one side of the funnel) systematically removed.

This paper describes a simple meta-analysis of the published detections of highly shifted, narrow X-ray lines in active galaxies using a funnel plot-like analysis. The conventional funnel plot would not be appropriate in the present context because there is expected to be considerable intrinsic heterogeneity in the strength and properties of the shifted lines, which means there is no single, true value for the strength of the ``effect.'' But the principle of the funnel plot should still hold: the estimated strengths of the lines should be independent of the quality of the data used to find them (in this context quality means essentially the signal-to-noise of the data, which is of course closely related to the size of the photon sample that constitutes the spectrum). The strength of a real line should be independent of the exposure time and detector sensitivity used to measure it. 


\section{Analysis}
\label{sect:analysis}

The starting point for the meta-analysis was a search for published claims of highly shifted, narrow, emission or absorption lines in the X-ray spectra of Seyfert galaxies and quasars. For the purposes of the present study, these features are defined as intrinsically narrow\footnote{Narrow in this context is defined to mean unresolved or marginally resolved in the data, and with $\sigma \lsim 0.1$~keV. The criterion was used to filter out broad absorption troughs, blends and photoelectric edges} emission or absorption features found in $1.5-20$~keV X-ray spectra of Seyfert galaxies or quasars, that have been identified with prominent transitions in the X-ray band (e.g. K$\alpha$ lines of Fe, Ca, Ar, S, Si, Mg) leading to inflow/outflow velocities $v/c \ge 0.05$. 
In the particular case of iron, emission lines were accepted if outside the range $6.1-7.3$ keV (corresponding to $6.4-6.9$ keV K$\alpha$ lines from Fe {\sc i - xxvi} at $v/c = \pm 0.05$), and absorption lines if outside the range $6.4-7.3$ keV (corresponding to $6.7-6.9$ keV resonances in H and He-like Fe {\sc xxv-xxvi}). The slightly different ranges for absorption and emission correspond to the different species expected to dominate in each case.
This is criterion provides a simple and uniform, albeit arbitrary, way to distinguish between the highly shifted, isolated, narrow line features that are the focus of this paper and more mildly shifted structures likely to be more directly linked to a broad emission (or absorption) complex centred around $6.4-6.9$ keV. The main result and conclusion of the paper would not be significantly changed if the $v/c$ threshold was increased (e.g. to  $v/c \ge 0.1$).

An initial list of papers was constructed from all refereed articles listed in the NASA Astrophysics Data System (ADS\footnote{\tt{http://adswww.harvard.edu/}}) published between 1995 and 2007 (inclusive), selecting papers with abstract text that matched the Boolean expression ``narrow {\it and} X-ray {\it and} line {\it and} (redshifted {\it or} blueshifted).'' The resulting $135$ papers were then examined individually to select only those that reported new detections of the type of features under investigation. This provided a list of $12$ such papers. By following their ``paper trails'' (citations to/from the articles) it was possible to add a further $14$ papers, yielding a total of $26$ papers presenting detections $38$ shifted, narrow lines. Table~\ref{table} lists all the line features found through this literature search. 

The X-ray absorption systems reported in the gravitationally lensed Broad Absorption Line (BAL) quasars \citep[][]{chartas02, chartas03, chartas07b, chartas07a} were treated separately. In every published BAL case at least one component of the absorption system was reported as resolved and broad, and so did not match the criterion above. Also, gravitationally lensed BAL quasars arguably represent a rather special sample of objects within which  there known high velocity absorption systems, and so there are good reasons to treat these objects as distinct from the sample of Seyferts and non-BAL quasars. For completeness these are included in Table~\ref{table}, but are not considered in the discussion that follows.

\begin{table*}
\caption{Data and sources used in the meta-analysis. The columns list the following information: (1) source name, (2) redshift, (3) the exposure time of the observation, (4) centroid energy, (5) equivalent width ($EW$) and (6) its 90 per cent uncertainty for the line detections, (7) the stated improvement in the fit statistic due to the line, and (8) the corresponding reference. Negative $EW$s indicate absorption lines. The `$^f$' symbol indicates the line strength was given only in photon flux ($10^{-5}$ ph s$^{-1}$ cm$^{-2}$) terms, but converted into $EW$ (see text). Column (3) also indicates the X-ray mission: X (\xmm), C (\chandra), A (\asca), B (\sax). The lensed BAL quasars (see text) are listed separately at the bottom of the table. The results from \citet{pounds05} replace those from \citet{pounds03a}.
}
\label{table}
\centering
\begin{tabular}{llrlrrrl}
\hline\hline
Target         &  $z$  &$T_{\rm exp}$& $E$ & $EW$ & $err(EW)$ & $\Delta \chi^2$& reference \\
name           &       & (ks)  &(keV)& (eV) & (eV) &  &  \\
(1)            & (2)   & (3)   & (4) & (5)  & (6)  & (7) & (8) \\
\hline
PKS 0637-75    & 0.653 & 49 (A)& 1.6 & -58  & 36   &14.6&  \cite{yaqoob98}\\ 
PG 1211+143    & 0.081 & 50 (X)& 1.63& -14  & 3    & 32 &  \cite{pounds03a,pounds05}\\ 
4U 1344-60     & 0.013 & 26 (X)& 1.63& 19   & 11   & -- &  \cite{piconcelli06}\\ 
PG 1211+143    & 0.081 & 50 (X)& 2.94& -36  & 9    & 26 &  \cite{pounds03a,pounds05}\\ 
PG 0844+349    & 0.064 & 20 (X)& 3.02& -35  & 16   & 7  &  \cite{pounds03b}, but see \citep{brinkmann06}\\ 
NGC 4151       & 0.003 & 69 (X)& 3.70& 10   & 5    &15.6& \cite{nandra07} \\ 
PG 1211+143    & 0.081 &134 (C)& 4.22& -35  & 16   &13.8&  \cite{reeves05}\\
Mrk 841        & 0.036 & 30 (X)& 4.80& 50   & 20   & 11 &  \cite{petrucci07}\\ 
4U 1344-60     & 0.013 & 26 (X)& 4.9 & 45   & 23   &--  &  \cite{piconcelli06}\\ 
PG 1211+143    & 0.081 &134 (C)& 4.93& -57  & 23   &19.9&  \cite{reeves05}\\ 
NGC 4151       & 0.003 & 32 (X)& 5.23& 8    & 4    &18.9& \cite{nandra07} \\ 
4U 1344-60     & 0.013 & 26 (X)& 5.3 & 57   & 28   & -- &  \cite{piconcelli06}\\ 
Q0056-363      & 0.162 & 94 (X)& 5.34& -75  & 37   & -- & \cite{matt05} \\ 
ESO 113-G010   & 0.026 & 4  (X)& 5.38& 265  & 90   &9.6 &  \cite{porquet04} \\   
Mrk 509        & 0.034 & 33 (B)& 5.45& -173 & 146  &8.7 &  \cite{dadina05} \\   
PG 1416-129    & 0.129 & 50 (X)& 5.5 & 194  & 89   &12.8& \cite{porquet07a} \\ 
Mrk 509        & 0.034 & 33 (B)& 5.5 & -195 & 83   & 16.8 & \cite{dadina05} \\ 
NGC 3516       & 0.009 & 75 (C)& 5.57& 23$^f$ & 4$^f$ & --   & \cite{turner02} \\
Mrk 766        & 0.013 &130 (X)& 5.60& 18$^f$ & 9$^f$ & 12   &  \cite{turner04}\\ 
ESO 198-G24    & 0.046 & 7  (X)& 5.7 & 70   & 40   & 9.3&  \cite{guainazzi03,bianchi04}\\ 
Mrk 766        & 0.013 & 130(X)& 5.75& 56$^f$ & 23$^f$ & 13   &  \cite{turner04}\\ 
NGC 7314       & 0.005 & 97 (C)& 5.84& 32   &  16  & -- &  \cite{yaqoob03} \\ 
NGC 3516       & 0.009 &152 (A)& 5.9 & $\sim$30   & --   &28.3& \cite{nandra99} \\
Mrk 335        & 0.026 & 30 (X)& 5.92& -50  & 21   & 16 & \cite{longi07} \\ 
Ark 120        & 0.033 & 57 (X)& 6.01& -21  & 10   &21.3& \cite{nandra07} \\
NGC 3227       & 0.004 &100 (X)& 6.04&  21  & 9    &24.9 & \cite{markowitz08}\\
NGC 3516       & 0.009 & 57 (X)& 6.08& $\sim$40 & --   & -- & \cite{bianchi04,dovciak04}\\
E1821+643      & 0.297 &100 (C)& 6.2 & -54  & 13   & -- &  \cite{yaqoob05}\\
NGC 4151       & 0.003 & 69 (X)& 7.33& -15  & 8    &16.8& \cite{nandra07} \\ 
NGC 4151       & 0.003 & 32 (X)& 7.45& -16  & 6    &28.0& \cite{nandra07} \\ 
RX J0136.9-3510& 0.289 &195 (A)& 7.6 & 860  & 401  &11.0&  \cite{ghosh04}\\ 
PG 1211+143    & 0.081 & 50 (X)& 7.61& -105 & 35   &32  &  \cite{pounds03a,pounds05}\\ 
IC 4329A       & 0.016 & 69 (X)& 7.68& -15  & 7    &21.0&  \cite{markowitz06,nandra07}\\ 
MCG-5-23-16    & 0.008 & 96 (X)& 7.7 & -33  & 10   & 20 & \cite{braito07}\\ 
UGC 3973/Mrk 79& 0.022 & 4  (X)& 7.99& 161  & 89   &8.1 &  \cite{gallo05} \\  
Mrk 509        & 0.034 & 33 (B)& 8.14& -383 & 150  &16.4&  \cite{dadina05} \\  
PG 0844+349    & 0.064 & 20 (X)& 8.18& -170 & 60  &11  &  \cite{pounds03b}, but see \citep{brinkmann06}\\ 
PKS 2149-306   & 2.345 & 20 (A)& 17.0& 298  & 204  &10.3&  \cite{yaqoob99}\\ 
\hline
\multicolumn{8}{c}{{\it Lensed BAL quasars}}\\
\hline
PG 1115+080    & 1.72  & 63 (X) &7.38&-140  & 60   &--  &  \cite{chartas03} \\
APM 08279+5255 & 3.91  & 89 (C)& 8.05&-240  & 70   &35.2&  \cite{chartas02} \\
H 1413+117     & 2.56  & 89 (C) & 8.5& --   & --   & -- &  \cite{chartas07a} \\
PG 1115+080    & 1.72  & 63 (X) &9.50&-1400 & 500  &--  &  \cite{chartas03} \\
APM 08279+5255 & 3.91  & 89 (C)& 9.79&-430  & 150  &40.2&  \cite{chartas02} \\
H 1413+117     & 2.56  & 89 (C) &13.9&-2400 &1600  & -- &  \cite{chartas07a} \\
\hline
\end{tabular}
\end{table*}

The strength of the feature reported by \citet{gallo05} was given in both equivalent width ($EW$) and photon flux terms but the relative uncertainties stated for each are different: $13$ and $55$ per cent, respectively. Given the modest effect of this feature on the fit statistic ($\Delta \chi^2 = 8.1$), the larger of these two uncertainties would appear to be the more plausible, and this value is used in Table~\ref{table}, although it should be noted that the conclusion of the present paper does not depend on this one value.
In the cases of Mrk 766 \citep{turner04} and NGC 3516 \citep{turner02}, the line strengths were given only in flux terms. In order to provide a better comparison with the other lines these were converted in $EW$ terms using the flux density of the continuum at the location of the lines, found by fitting the relevant data\footnote{The \xmm\ data for Mrk 766 were obtained through the XSA, processed with SAS v7.1.0, and EPIC pn spectra for ``high'' and ``low'' periods were extracted from the first $100$ ks and last $\sim 30$ ks, as described by \citet{turner04}. The $3-11$ keV spectra were then fitted with a power law model, excluding the $5-7$ keV band \citep[following][]{turner04}. The \chandra\ HETGS data for NGC 3516 were obtained through the HotGAS database at {\tt http://hotgas.pha.jhu.edu/} and the HEG ($3-9$~keV) and MEG ($3-6.5$~keV) spectra fitted simultaneously with a power law model after excluding the $5-7$ keV interval.}. The $5.9$ keV absorption line in NGC 3516 has no published $EW$ \citep{nandra99}, although spectral fitting of the publically available data yielded an estimate of $\sim -30$ eV\footnote{The data were obtained from the Tartarus database at {\tt http://astro.ic.ac.uk/research/tartarus/} and fitted over the $3-10$~keV range with a power law plus Laor diskline model, and a Gaussian absorption line, as discussed in \citet{nandra99}.}. The $EW$ of the $6.08$~keV emission line in NGC 3516 was not stated by \citet{bianchi04} but the analysis of \citet{dovciak04} would seem to suggest it is about $\sim 40$~eV. Both the $5.9$ keV and $6.08$ keV lines in NGC 3516 were published without estimates of the uncertainty on their strengths.

\section{Results}

\begin{figure*}
\begin{center}
   \includegraphics[width=11.0 cm, angle=90]{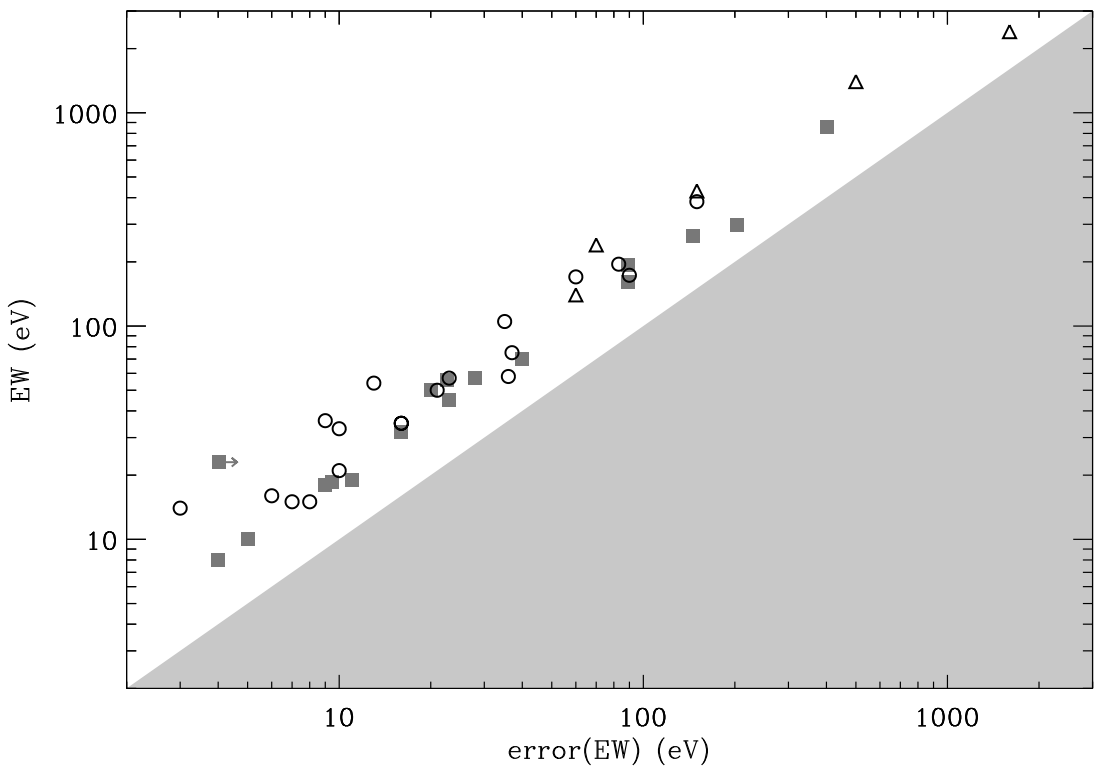}
\end{center}
\caption{
The ``signal-error'' diagram for narrow, relativistically shifted iron lines compiled from the data given in Table~\ref{table}. The diagram shows the estimated line strengths (absolute value of $EW$) against the uncertainty in the estimate (half width of the stated 90 per cert confidence interval). The shaded region indicates the zone of avoidance within which a feature would not be reported as ``detected,'' i.e. $EW \lsim err(EW)$. Open circles indicate absorption lines, filled squares indicate emission lines, and open triangles represent absorption systems from lensed BAL quasars (see text). The $5.6$ keV emission line in NGC 3516 is marked with an arrow to indicate its systematically underestimated uncertainty (see text). 
\label{fig}
}
\end{figure*}

Excluding the lensed BAL quasars, the compiled data list $36$ lines from $23$ sources with $EW$s and $90$ per cent uncertainties\footnote{As far as can be ascertained, all the confidence regions were calculated by varying the $EW$ parameter until the observed fit statistic ($\chi^2$ or $C$-statistic) increased by $2.706$ over its minimum. In cases where the confidence interval was roughly symmetric about the best fit, the half width of the stated $90$ per cent confidence region was taken as a single estimate of the uncertainty. In the case of highly asymmetric intervals, the part of the interval extending below the best fit (i.e. towards zero $EW$) was used.} which are indicators of the ``signal'' and ``noise'' respectively (similar to the ``effect strength'' and ``sample size'' of the funnel plot). Of these, $17$ are for emission lines and $19$ from absorption lines. 
Figure~\ref{fig} shows a scatter diagram for these two quantities. This diagram serves the same purpose as the funnel plot, showing whether the strength of the measured effect ($EW$) depends on the quality of the data (as indicated by the uncertainty on $EW$). If most of the lines are real, some observations should populate the upper left portion of the diagram. The zone of avoidance in the lower right arises from the fact that any line with a $90$ per cent confidence interval on its $EW$ that includes (or extends very close to) zero, would probably not be reported as a detection. Despite two decades of range in the line strengths there is a clear trend for all the data points to lie close to the edge of the zone of avoidance, i.e. just above the detection limit, irrespective of the line strength. If the lines do indeed span this range in strengths, the strongest lines should be easily detectable in the best observations (i.e. those with smallest uncertainties), but only weak features are claimed in all these cases. It would appear that the strength of any narrow, relativistically shifted lines depends on the quality of the data they were detected in.

\section{Discussion}
\label{sect:disco}

The tendency for stronger lines to be accompanied by proportionately larger uncertainties (as shown in Figure~\ref{fig}), or equivalently the relatively constant $|EW|/error$ ratio over the large range in $|EW|$, requires explanation. The lines with larger (absolute) $EW$ should be easy to detect in more sensitive observations (and should give smaller uncertainties) and so should populate the upper-left region of Figure~\ref{fig}, but this is not the case. One simple explanation is that many of the line detections are the most `significant' false detections from a large population of data covering a wide range of power to detect potential lines. 
But this begs the question of why are there so many false detections? 

Examination of a spectrum in isolation may or may not provide a false detection, but an apparently stronger detection will be more strongly favoured for publication. The many hundreds of observations that have been examined and not provided detections (whether with strong or weak limits) would populate the shaded region of Fig.~\ref{fig} if only they were published (very few observations are published with limits on the strength of undetected narrow, shifted lines). The reported detections may be in effect the ``tip of the iceberg'' -- the strongest or most significant of a population of random fluctuations, with the rest of the population unseen due to publication bias. If the lines are genuine, the challenge is to explain why all the detections are close to the detection limit despite the huge range in the quality of the data, or equivalently, why the largest (absolute) $EW$ lines appear only in the poorest data with the largest error bars. Of course it is plausible that Fig.~\ref{fig} shows a mixture of false and true detections, with the true detections expected mostly among the weaker lines (e.g. $|EW| < 30$~eV) since they are necessarily limited to moderately weak detections even in the best data (which is not true of stronger lines).

\subsection{Detection methods}
\label{sect:detect}

Virtually all these lines were justified on the basis of $p$-values from hypothesis tests being smaller than some threshold $\alpha$, a process intended to limit the fraction of false detections (Type I errors) to $\alpha$, and in almost all the cases listed above a reasonable detection significance level (e.g. $\alpha = 0.05$ or $0.01$) was used. Perhaps the $p$-values were systematically underestimated, leading to an abundance of false detections? 

Estimating the statistical significance of a possible line feature is indeed rather difficult \citep[see][]{freeman99,protassov02}, and many authors have used and continue to use inappropriate tests (e.g. the $F$-test) which may inflate the number of false detections. 
Ideally, a thorough and uniform analysis of all the data would solve the problem that different data analysis techniques were used by different authors, but that is not the purpose of the present paper, which simply makes use of the published results as they are presented in the literature. However, even a uniform and comprehensive analysis of all the datasets listed in Table~\ref{table}, using sophisticated statistical methods, would not change the basic result (the clustering of lines near the diagonal in Figure~\ref{fig}) unless it was true that many of the uncertainties listed in the table are greatly exaggerated. If the re-calculated uncertainties were often much smaller the typical $|EW|/error$ would be much higher, and the tendency for strong lines to have proportionately larger uncertainties might be eroded -- the points would fill more of the top-left region of the figure rather than skirting the diagonal. But there is no clear reason to suspect this might be true. If a re-analysis found the claimed significances (e.g. the $p$-values used for detection) were slightly too high or too low this may affect the number of points in the figure, but not the trend it reveals. If the $p$-values were systematically far too small (i.e. too significant) there must be a large number of false detections present in the figure, whereas if the $p$-values are reliable (or even overestimated) the $|EW| - error$ relation still requires an explanation. 
Therefore, in the remainder of this discussion it is assumed that the statistical tests used in the papers listed in Table~\ref{table} are sound, and that it is still necessary to seek an explanation for the relation shown in Figure~\ref{fig}.

\subsection{Confounding factors}

The observed effect could be produced from genuine line detections if the two variables, line strength and its uncertainty (i.e. data quality), were correlated with some other factor, such as source distance. Perhaps the more distant, luminous quasars, that often yield the poorest spectra, possess intrinsically stronger line features compared to the nearby, low-luminosity Seyferts that have high-quality spectra available. Indeed, there is a correlation between redshift and $EW$ (Spearman rank-correlation coefficient $\rho = 0.62$, $p = 5.6 \times 10^{-5}$), but it is much weaker than between $|EW|$ and its uncertainty ($\rho = 0.96$, $p < 10^{-15}$), and is due  entirely to the four lowest redshift sources (NGC 4151, NGC 7314, NGC 3516 and NGC 3227 at $z \le 0.01$), which all have long exposures and showed only very weak lines. If these sources are ignored (leaving $28$ lines) there is no significant correlation between redshift and $|EW|$ ($\rho = 0.33$, $p=9.0\times 10^{-2}$) or its uncertainty ($\rho = 0.24$, $p = 0.22$). The tight, linear relation between $|EW|$ and its uncertainty involves all sources and so cannot be due to the $EW$ (or its uncertainty) being correlated with redshift. Also, as can be seen from Table~\ref{table}, the strong lines from ESO 113-G010, ESO 198-G24 and Mrk 79 have large errors because the observations were very short ($<8$~ks), not because the targets are intrinsically faint. In principle longer observations of these strong lines should conclusively demonstrate the reality of these features. However, a subsequent, much longer ($100$~ks) observation of ESO 113-G010 by \citet{porquet07b} did not detect the $5.4$ keV emission line previously found in a $\sim 4$ ks exposure \citep{porquet04}. If real, the $EW$ must have decreased tenfold between observations to avoid detection by \citet{porquet07b}; either a coincidence or an indication that the original detection of a strong line was false. 

Another possible confounding factor is line energy. The continuum spectrum of the active galaxies listed in Table~\ref{table} is usually well described by a power law with a photon index typically in the range $\Gamma \sim 1.5 - 2.5$, meaning there are far fewer photons at higher energies than lower energies. If narrow lines appearing at higher energies (e.g. blueshifted Fe K$\alpha$ at $E \gsim 7.1$~keV) tended to have higher $|EW|$ than those at lower energies (e.g. redshifted Fe K$\alpha$ at $E \lsim 6.2$~keV), the strongest lines may have the  largest uncertainties simply because they occur preferentially at higher energies. There is very little correlation between line energy and $|EW|$ ($\rho = 0.27$, $p=0.11$) or its uncertainty ($\rho = 0.26$, $p = 0.13$). Indeed, Table~\ref{table} shows that some of the smallest $EW$s (and uncertainties) were found in lines above $7.1$~keV, and the tight relation between strength and uncertainty is present in the subset of higher-energy lines. It is not clear what other source property could be the confounding factor needed to explain the absence of strong lines in the best observations.

There is one other point worth emphasizing about the line energies. Of the $38$ lines listed in Table~\ref{table}, $14$ were found in the range $5-6$ keV. The excess of detections could indicate that modestly redshifted ($z \sim 0.05 - 0.3$) emission or absorption are more significant or robust than the more strongly shifted features, i.e. the number of detections is enhanced because there are more `true' lines in this band. However, the findings of this paper remain unchanged if one considers only the $13$ lines in the $5-6$ keV range with errors: these lines follow the same $|EW| - error$ correlation as shown in Fig~\ref{fig}, and the $|EW|/error$ is not higher for this subsample. Indeed, the mean $|EW|/error = 2.4$ for the $5-6$ keV subset, and $2.5$ for the remaining lines. (Excluding the $5.57$ keV line in NGC 3516, for reasons discussed below, the $5-6$ keV subsample gives a mean $|EW|/error = 2.1$.) Even for this subsample the problem remains to explain the tight, linear correlation between $|EW|$ and its uncertainty. There are at least two possible effects that might help explain an enhanced rate of false detections in this narrow band. The first is confusion with emission structure from a strong, broad (possibly asymmetric) line centred around 6.4 keV. Such emission may produce an excess of counts in the $5-6$ keV region, above the expected continuum, and so enhance the appearance of line-like residuals. The second effect is observer bias; it is plausible that individual observers preferentially attend to residuals in the $5-7$ keV region immediately around expected Fe structures. Very few of the papers listed in Table~\ref{table} report that the detections were made on the basis of a systematic and uniform search over a wide spectral range (a notable exception is \citet{nandra07}). Either or both these effects may enhance the number of false detections made in the $5-6$ keV range. 

\subsection{The strongest individual cases}

Of the $36$ lines listed in Table~\ref{table} (ignoring the BALs) with strength and uncertainty estimates, only five have strengths that are greater than three times their uncertainty (half width of the 90 per cent confidence interval). These are: emission at $5.57$ keV in NGC 3516 \citep[$EW \approx 23$~eV;][]{turner02}, absorption at $6.2$ keV in E1821+643 \citep[$EW \approx -54$~eV;][]{yaqoob05}, absorption at $7.7$ keV in MCG-5-23-16 \citep[$EW \approx -33$~eV;][]{braito07}, $1.63$ keV and $2.94$ keV absorption in PG 1211+143 \citep[$EW \approx -14$~eV and $-36$~eV;][]{pounds03a,pounds05}. These are considered in turn below.

The emission line in NGC 3516 was found in \chandra\ HETGS data, but could not be detected in the partly simultaneous \xmm\ data, those data gave an upper limit on the line flux an order of magnitude smaller than the \chandra\ detection \citep{turner02}. If real, the line flux must have decreased by at least one order of magnitude between the observations. Furthermore it should be noted that the line energy was held fixed during the evaluation of the confidence interval on the flux \citep[see the footnote to Table~1 of][]{turner02}. 
As the line energy was not predicted but obtained from fitting the data, it should have remained a free parameter throughout the calculation of confidence intervals, otherwise the confidence interval may be artificially reduced. In any event, the procedure described by \citet{turner02} differs from the standard procedure adopted in the other cases and so is perhaps best considered a lower limit on the size of the confidence region of that particular line (as indicated in Figure~\ref{fig}). 

The absorption line found in the \chandra\ observation of E1821+642 was seen in the HEG spectrum but could not be confirmed in the lower signal-to-noise MEG data. In the case of PG 1211+143 the simultaneous detection of multiple lines, and their identification at similar blueshifts, in both CCD (EPIC) and grating (RGS) data from \xmm\ \citep{pounds03a} would seem to put this case on firmer ground. For completeness, it should be noted that the identification of the lines in terms of highly blueshifted features has been debated, e.g.
 \citet{mckernan05,kaspi06,pounds07,reeves08}. \citet{reeves05} presented \chandra\ grating data of the same object and again reported absorption lines, except redshifted not blueshifted.
The $7.6$~keV absorption line found in the \xmm\ data was not detected in more recent \suzaku\ data \citep[with an upper limit $\sim 4$ times smaller than the original \xmm\ detection;][]{reeves08}. 

\citet{braito07} and \citet{reeves07} studied MCG-5-23-16 using simultaneous \xmm, \chandra, \suzaku\ and \xte\ observations. The detection of $7.7$ keV absorption is based on the EPIC pn spectrum from \xmm. The EPIC MOS spectrum is consistent with the pn spectrum but is unable to confirm the presence of the line due to the smaller photon sample size; the \suzaku\ data show a possible absorption feature, but it was poorly constrained compared to the EPIC pn data; and the \chandra\ data were unable to confirm the line detection.

The lines that gave the largest improvement in the $\chi^2$ fit statistic were both absorption lines: at $5.9$~keV in NGC 3516 (with \asca) and at $7.45$~keV in NGC 4151 (with \xmm). The former has no published EW, while the latter has a surprisingly low $|EW|/error$ ratio given its apparent effect on the fit. However, as \citet{nandra07} noted, there is some ambiguity over whether the NGC 4151 feature should be identified with a line or an edge (see their section 8.8.4).

The Seyfert galaxy NGC 3516 provides one more interesting example. \citet{dovciak04} described a $6.08$~keV emission line in an \xmm\ EPIC pn spectrum of NGC 3516 taken in 2001 April \citep[see also][]{bianchi04} and suggested the feature is varied in $EW$ and/or energy. The significance of the feature was assessed using an $F$-test (see section~\ref{sect:detect}) but no uncertainties were given on the $EW$ and so the observation is not represented in Figure~\ref{fig}. The same observation was analysed by \cite{iwasawa04}, who claimed a periodic modulation in the spectral shape of a broad $5.6-6.5$ keV line based on $\sim 3$ `cycles'. Although this claim is intriguing it is not an independent confirmation of the significance of the $\sim 6.1$~keV line; the analysis is an attempt to better understand and model the feature reported by \citet{dovciak04}, assuming its reality and using the same data, not an independent assessment of it.

\subsection{The effect of selection and publication bias}

This leaves the possibility that many or most of the line detections are false detections caused by random sampling fluctuations. The number of false detections may at first sight appear large, but one must remember that each spectrum from \xmm, \chandra\ etc. contains $\gsim 50$ resolution elements, and there are many hundreds of spectra (especially considering that longer observations are routinely split into multiple spectra corresponding to different time intervals, flux levels, etc.). The non-detections from each resolution element of each spectrum contributes an (unpublished) point inside (or just above) the shaded region of Fig.~\ref{fig}. The human analyst, or an automated search algorithm, has a tendency to focus on the largest fluctuations -- this is a selection bias. These may then be subjected to an hypothesis test, and those satisfying some conventional criterion ($p < \alpha$ with e.g. $\alpha = 0.05$) may be chosen for publication. The more `significant' the result (i.e. the smaller $p$ is), the more likely it is to be chosen for publication: publication bias. These two biases act in the same direction but at different stages in the process, and are examples of what Francis Bacon, in his \emph{Novum Organum}, described as the tendency to ``notice the events where they are fulfilled, but where they fail, though this happens much more often, neglect and pass them by'' \citep{bacon1620}. The result is that the vastly greater number of null results (whether with strong or weak limits) will go largely unpublished, making it difficult to estimate the global significance of any individual detection. 

The question that needs to be addressed is whether any given excess or deficit in a spectrum is unlikely to be a sampling fluctuation given a large number of spectra each with many resolution elements. The only systematic attempts to address this specific problem are those of \citet{nandra07} and \citet{longi06}, both of which describe surveys of narrow, shifted lines in samples of observations.

One can make an order of magnitude estimate of the number of unpublished non-detections using the following simple argument. Let us assume that $\gsim 500$ spectra have been examined in the last few years and each has $\gsim 50$ resolution elements, the number of independent spectral resolution elements that have been examined must be $\gsim 2.5 \times 10^4$. If the residuals (after fitting a suitable continuum model) in each of these is approximately Normally distributed, the expected numbers of fluctuations at $|z| \ge 2\sigma$, $3\sigma$ and $4 \sigma$ are $\gsim 1137.5$, $\gsim 67.5$ and $\gsim 1.6$. The data in Table~\ref{table} record $23$ detections with $|EW|/error > 2$ and $5$ with $|EW|/error > 3$, given that the quoted errors are the half widths of the $90$ per cent confidence interval, these might better correspond to $\sim 3\sigma$ and $\sim 4\sigma$ detections\footnote{This is a crude approximation. Interval estimation and hypothesis testing are different statistical procedures and the relative uncertainty of the strength of a line will not in general be simply related to its significance in an hypothesis test.}, suggesting the expected number of large fluctuations is consistent with the amount of available data. Furthermore, it seems reasonable that lower significance features are less likely to be reported, and therefore  the fraction of unreported $\sim 3\sigma$ detection would be larger than the fraction of unreported $\sim 4\sigma$ detections, hence the lower than expected number of (reported) detections with $|EW|/error$ in the range $2-3$. Another way to consider this is to compare the significance of individual lines, estimated accounting for the number of spectral channels, to the number of examined spectra. For example, one of the best detections is in E1821+643, where \citet{yaqoob05} estimate a $2-3\sigma$ detection over the entire spectrum, but given $\gsim 500$ spectra one would expect $\gsim 22.7$ detections at better than $2 \sigma$ and $\gsim 1.3$ at better than $3 \sigma$, so perhaps this is not unexpected. See \citet{scargle00} for a discussion of methods to estimate the number of unpublished observations.

\subsection{\emph{Post hoc} and \emph{a priori} arguments}

One cannot argue \emph{post hoc} that there are particular properties of individual datasets that allow them to be considered in isolation\footnote{The possible exceptions are the lensed BAL quasars, which are exceptional sources for which there is prior knowledge of high velocity outflows (from their rest-frame ultraviolet spectra). One might expect stronger absorption systems from these faint sources, and so they might reasonably be considered separately (they were not included in the calculations performed above).}. For example, in the case of E1821+643 one might argue that the $6.2$~keV line was detected in high resolution \chandra\ HETGS data and there are far fewer of these observations, therefore the number of (unpublished) non-detections is much smaller. One could conceivably construe a similar argument in favour of the uniqueness of almost any observation. But if $N$ observations are analysed and subjected to hypothesis tests with a detection threshold of $\alpha$ (it does not matter if the tests are the same), the expected number of false detections is $\sim N \alpha$, even if the data come from different target sources, missions, detectors etc. 

Similarly, it is not valid to argue that certain time intervals of a particular observation should be treated as special unless they are selected on the basis of an explicit, prior criterion (i.e. derived and used independently of line detection). The expected number of false detections scales with the number of tests performed, therefore if a long, high quality observation is split into ten time intervals and each is examined (if only in a cursory fashion), this has increased the effective number of tests approximately tenfold. Additionally, one cannot engineer the data slicing to maximise the detection of a line, based on the detection of the line in the same data finely sliced, and consider this a fair test. This would be to test a hypothesis suggested by the data, as if the data were independent; one may equally well roll a die fifty times, find the most frequent number to be rolled and then claim the die is biased because there is only a $1/6$ chance of that number being the most frequent.

The plausibility of a line detection would be augmented if there was cogent prior information on the line properties that was confirmed in subsequent observations. For example, a small excess in a spectrum at $6.4$~keV might be considered a significant detection of an iron line, but the same excess at an arbitrary energy might not be significant. The coincidence of an excess appearing at the predicted energy adds to its plausibility as a real line (and this can be included in formal calculations of its `significance'). 
However, as discussed above, the lines appear at arbitrary energies and are often reported to be transient which, if true, makes prediction difficult. 

If real, these lines represent quite extreme physical phenomena. But, at least in the case of absorption lines, low outflow-velocity absorption systems are routinely seen in the X-ray spectra of Seyfert galaxies and quasars, and high velocity outflows are observed in the rest-frame ultraviolet spectra of BAL quasars. Based on these one may argue that the existence of higher velocity X-ray outflows in other (i.e. non-BAL) sources is at least plausible. By contrast the existence of narrow, redshifted absorption, and highly red- or blue-shifted emission lines is unprecedented, and it is reasonable to demand high standards of evidence to support their existence. It is perhaps worth noting that the $15$ emission lines from Table~\ref{table} have a lower average $|EW|/error$ ratio than the absorption lines, and none of them has $|EW|/error > 2.5$ (the $5.6$~keV emission line in NGC 3516 is excepted for the reasons given above).

\subsection{Final remarks}

Of course, new and exciting discoveries are usually made at the limits of the available data, but these must be confirmed at higher significance by subsequent observations; when more detections are made but the significance does not improve despite more, longer observations and brighter targets one should not automatically consider the discovery confirmed. It is not the objective of this paper to argue that any specific detection is false; the argument based on Fig.~\ref{fig} is purely statistical. Indeed, several of the existing detections may be genuine -- the case is arguably strongest for absorption lines with the smallest $EW$ (e.g $EW \lsim 30$~eV) and largest $\Delta \chi^2$ -- but it is difficult to explain the tightness of the correlation between detected line strength and its uncertainty if all or most of the detections are genuine. The prevalence and importance of such features in the population at large therefore remains an open question.


\section*{acknowledgements}
We thank an anonymous referee for a useful discussion on certain aspects of this paper.
PU acknowledges support from an STFC Advanced Fellowship. 
This research has made use of NASA's Astrophysics Data System. 
We would also like to thank Davide Lazzati, who first demonstrated to us the value of the signal-error plot (albeit in a different context). 
This research has made use of the HotGAS database, created and maintained under the supervision of Tahir Yaqoob. HotGAS has been supported by a \chandra\ archival research grant, AR4-5009X, issued by the \chandra\ X-ray Observatory center, which is operated by the Smithsonian Astrophysical Observatory for and on behalf of NASA under contract NAS8-39073.
This research has made use of the Tartarus (Version 3.2) database, created by Paul O'Neill and Kirpal Nandra at Imperial College London, and Jane Turner at NASA/GSFC. Tartarus is supported by funding from STFC, and NASA grants NAG5-7385 and NAG5-7067.


\bibliographystyle{mn2e}
\bibliography{mnras_refs}

\begin{thebibliography}{53}
\expandafter\ifx\csname natexlab\endcsname\relax\def\natexlab#1{#1}\fi

\bibitem[{{Ariew} \& {Watkins}(2000)}]{bacon1620}
{Ariew} R., {Watkins} E., eds., 2000, {Readings in Mordern Philosophy Volume
  I}. Hackett Publishing Co.

\bibitem[{{Begg} \& {Berlin}(1988)}]{begg88}
{Begg} C.~B., {Berlin} J.~A., 1988, J. R. Stat. Soc. A, 151, 419

\bibitem[{{Bianchi} {et~al.}(2004){Bianchi}, {Matt}, {Balestra}, {Guainazzi},
  \& {Perola}}]{bianchi04}
{Bianchi} S., {Matt} G., {Balestra} I., {Guainazzi} M., {Perola} G.~C., 2004,
  \aap, 422, 65

\bibitem[{{Braito} {et~al.}(2007){Braito}, {Reeves}, {Dewangan}, {George},
  {Griffiths}, {Markowitz}, {Nandra}, {Porquet}, {Ptak}, {Turner}, {Yaqoob}, \&
  {Weaver}}]{braito07}
{Braito} V., {Reeves} J.~N., {Dewangan} G.~C., {George} I., {Griffiths} R.~E.,
  {Markowitz} A., {Nandra} K., {Porquet} D., {Ptak} A., {Turner} T.~J.,
  {Yaqoob} T., {Weaver} K., 2007, \apj, 670, 978

\bibitem[{{Brinkmann} {et~al.}(2006){Brinkmann}, {Wang}, {Grupe}, \&
  {Raeth}}]{brinkmann06}
{Brinkmann} W., {Wang} T., {Grupe} D., {Raeth} C., 2006, \aap, 450, 925

\bibitem[{{Cappi}(2006)}]{cappi06}
{Cappi} M., 2006, Astronomische Nachrichten, 327, 1012

\bibitem[{{Chartas} {et~al.}(2003){Chartas}, {Brandt}, \&
  {Gallagher}}]{chartas03}
{Chartas} G., {Brandt} W.~N., {Gallagher} S.~C., 2003, \apj, 595, 85

\bibitem[{{Chartas} {et~al.}(2002){Chartas}, {Brandt}, {Gallagher}, \&
  {Garmire}}]{chartas02}
{Chartas} G., {Brandt} W.~N., {Gallagher} S.~C., {Garmire} G.~P., 2002, \apj,
  579, 169

\bibitem[{{Chartas} {et~al.}(2007a){Chartas}, {Brandt}, {Gallagher}, \&
  {Proga}}]{chartas07b}
{Chartas} G., {Brandt} W.~N., {Gallagher} S.~C., {Proga} D., 2007a, \aj, 133,
  1849

\bibitem[{{Chartas} {et~al.}(2007b){Chartas}, {Eracleous}, {Dai}, {Agol}, \&
  {Gallagher}}]{chartas07a}
{Chartas} G., {Eracleous} M., {Dai} X., {Agol} E., {Gallagher} S., 2007b, \apj,
  661, 678

\bibitem[{{Dadina} {et~al.}(2005){Dadina}, {Cappi}, {Malaguti}, {Ponti}, \& {de
  Rosa}}]{dadina05}
{Dadina} M., {Cappi} M., {Malaguti} G., {Ponti} G., {de Rosa} A., 2005, \aap,
  442, 461

\bibitem[{{Dov{\v c}iak} {et~al.}(2004){Dov{\v c}iak}, {Bianchi}, {Guainazzi},
  {Karas}, \& {Matt}}]{dovciak04}
{Dov{\v c}iak} M., {Bianchi} S., {Guainazzi} M., {Karas} V., {Matt} G., 2004,
  \mnras, 350, 745

\bibitem[{{Egger} {et~al.}(1997){Egger}, {Smith}, {Schneider}, \&
  {Minder}}]{egger97}
{Egger} M., {Smith} G.~D., {Schneider} M., {Minder} C., 1997, BMJ, 315, 629

\bibitem[{{Fabian} {et~al.}(2000){Fabian}, {Iwasawa}, {Reynolds}, \&
  {Young}}]{fabian00}
{Fabian} A.~C., {Iwasawa} K., {Reynolds} C.~S., {Young} A.~J., 2000, \pasp,
  112, 1145

\bibitem[{{Freeman} {et~al.}(1999){Freeman}, {Graziani}, {Lamb}, {Loredo},
  {Fenimore}, {Murakami}, \& {Yoshida}}]{freeman99}
{Freeman} P.~E., {Graziani} C., {Lamb} D.~Q., {Loredo} T.~J., {Fenimore} E.~E.,
  {Murakami} T., {Yoshida} A., 1999, \apj, 524, 753

\bibitem[{{Gallo} {et~al.}(2005){Gallo}, {Fabian}, {Boller}, \&
  {Pietsch}}]{gallo05}
{Gallo} L.~C., {Fabian} A.~C., {Boller} T., {Pietsch} W., 2005, \mnras, 363, 64

\bibitem[{{Ghosh} {et~al.}(2004){Ghosh}, {Swartz}, {Tennant}, {Wu}, \&
  {Ramsey}}]{ghosh04}
{Ghosh} K.~K., {Swartz} D.~A., {Tennant} A.~F., {Wu} K., {Ramsey} B.~D., 2004,
  \apjl, 607, L111

\bibitem[{{Guainazzi}(2003)}]{guainazzi03}
{Guainazzi} M., 2003, \aap, 401, 903

\bibitem[{{Iwasawa} {et~al.}(2004){Iwasawa}, {Miniutti}, \&
  {Fabian}}]{iwasawa04}
{Iwasawa} K., {Miniutti} G., {Fabian} A.~C., 2004, \mnras, 355, 1073

\bibitem[{{Kaspi} \& {Behar}(2006)}]{kaspi06}
{Kaspi} S., {Behar} E., 2006, \apj, 636, 674

\bibitem[{{Longinotti} {et~al.}(2006){Longinotti}, {Bianchi}, {Guainazzi},
  {Roa}, \& {Santos-Lleo}}]{longi06}
{Longinotti} A.~L., {Bianchi} S., {Guainazzi} M., {Roa} J., {Santos-Lleo} M.,
  2006, AN, 327, 1020

\bibitem[{{Longinotti} {et~al.}(2007){Longinotti}, {Sim}, {Nandra}, \&
  {Cappi}}]{longi07}
{Longinotti} A.~L., {Sim} S.~A., {Nandra} K., {Cappi} M., 2007, \mnras, 374,
  237

\bibitem[{{Markowitz} \& {et al.}(2008)}]{markowitz08}
{Markowitz} A., {et al.}, 2008, \apj, {submitted}

\bibitem[{{Markowitz} {et~al.}(2006){Markowitz}, {Reeves}, \&
  {Braito}}]{markowitz06}
{Markowitz} A., {Reeves} J.~N., {Braito} V., 2006, \apj, 646, 783

\bibitem[{{Matt} {et~al.}(2005){Matt}, {Porquet}, {Bianchi}, {Falocco},
  {Maiolino}, {Reeves}, \& {Zappacosta}}]{matt05}
{Matt} G., {Porquet} D., {Bianchi} S., {Falocco} S., {Maiolino} R., {Reeves}
  J.~N., {Zappacosta} L., 2005, \aap, 435, 857

\bibitem[{{McKernan} {et~al.}(2005){McKernan}, {Yaqoob}, \&
  {Reynolds}}]{mckernan05}
{McKernan} B., {Yaqoob} T., {Reynolds} C.~S., 2005, \mnras, 361, 1337

\bibitem[{{Nandra} {et~al.}(1999){Nandra}, {George}, {Mushotzky}, {Turner}, \&
  {Yaqoob}}]{nandra99}
{Nandra} K., {George} I.~M., {Mushotzky} R.~F., {Turner} T.~J., {Yaqoob} T.,
  1999, \apjl, 523, L17

\bibitem[{{Nandra} {et~al.}(2007){Nandra}, {O'Neill}, {George}, \&
  {Reeves}}]{nandra07}
{Nandra} K., {O'Neill} P.~M., {George} I.~M., {Reeves} J.~N., 2007, \mnras,
  382, 194

\bibitem[{{Naylor}(1997)}]{naylor97}
{Naylor} C., 1997, BMJ, 315, 617

\bibitem[{{Petrucci} {et~al.}(2007){Petrucci}, {Ponti}, {Matt}, {Longinotti},
  {Malzac}, {Mouchet}, {Boisson}, {Maraschi}, {Nandra}, \&
  {Ferrando}}]{petrucci07}
{Petrucci} P.~O., {Ponti} G., {Matt} G., {Longinotti} A.~L., {Malzac} J.,
  {Mouchet} M., {Boisson} C., {Maraschi} L., {Nandra} K., {Ferrando} P., 2007,
  \aap, 470, 889

\bibitem[{{Piconcelli} {et~al.}(2006){Piconcelli}, {S{\'a}nchez-Portal},
  {Guainazzi}, {Martocchia}, {Motch}, {Schr{\"o}der}, {Bianchi},
  {Jim{\'e}nez-Bail{\'o}n}, \& {Matt}}]{piconcelli06}
{Piconcelli} E., {S{\'a}nchez-Portal} M., {Guainazzi} M., {Martocchia} A.,
  {Motch} C., {Schr{\"o}der} A.~C., {Bianchi} S., {Jim{\'e}nez-Bail{\'o}n} E.,
  {Matt} G., 2006, \aap, 453, 839

\bibitem[{{Porquet} {et~al.}(2007{\natexlab{a}}){Porquet}, {Reeves},
  {Markowitz}, {Turner}, {Miller}, \& {Nandra}}]{porquet07a}
{Porquet} D., {Reeves} J.~N., {Markowitz} A., {Turner} T.~J., {Miller} L.,
  {Nandra} K., 2007{\natexlab{a}}, \aap, 466, 23

\bibitem[{{Porquet} {et~al.}(2004){Porquet}, {Reeves}, {Uttley}, \&
  {Turner}}]{porquet04}
{Porquet} D., {Reeves} J.~N., {Uttley} P., {Turner} T.~J., 2004, \aap, 427, 101

\bibitem[{{Porquet} {et~al.}(2007{\natexlab{b}}){Porquet}, {Uttley}, {Reeves},
  {Markowitz}, {Bianchi}, {Grosso}, {Miller}, {Deluit}, \&
  {George}}]{porquet07b}
{Porquet} D., {Uttley} P., {Reeves} J.~N., {Markowitz} A., {Bianchi} S.,
  {Grosso} N., {Miller} L., {Deluit} S., {George} I.~M., 2007{\natexlab{b}},
  \aap, 473, 67

\bibitem[{{Pounds} {et~al.}(2003b){Pounds}, {King}, {Page}, \&
  {O'Brien}}]{pounds03b}
{Pounds} K.~A., {King} A.~R., {Page} K.~L., {O'Brien} P.~T., 2003b, \mnras,
  346, 1025

\bibitem[{{Pounds} \& {Reeves}(2007)}]{pounds07}
{Pounds} K.~A., {Reeves} J.~N., 2007, \mnras, 374, 823

\bibitem[{{Pounds} {et~al.}(2003a){Pounds}, {Reeves}, {King}, {Page},
  {O'Brien}, \& {Turner}}]{pounds03a}
{Pounds} K.~A., {Reeves} J.~N., {King} A.~R., {Page} K.~L., {O'Brien} P.~T.,
  {Turner} M.~J.~L., 2003a, \mnras, 345, 705

\bibitem[{{Pounds} {et~al.}(2005){Pounds}, {Reeves}, {King}, {Page}, {O'Brien},
  \& {Turner}}]{pounds05}
---, 2005, \mnras, 356, 1599

\bibitem[{{Protassov} {et~al.}(2002){Protassov}, {van Dyk}, {Connors},
  {Kashyap}, \& {Siemiginowska}}]{protassov02}
{Protassov} R., {van Dyk} D.~A., {Connors} A., {Kashyap} V.~L., {Siemiginowska}
  A., 2002, \apj, 571, 545

\bibitem[{{Reeves} {et~al.}(2008){Reeves}, {Done}, {Pounds}, {Terashima},
  {Hayashida}, {Anabuki}, {Uchino}, \& {Turner}}]{reeves08}
{Reeves} J., {Done} C., {Pounds} K., {Terashima} Y., {Hayashida} K., {Anabuki}
  N., {Uchino} M., {Turner} M., 2008, \mnras, L14

\bibitem[{{Reeves} {et~al.}(2007){Reeves}, {Awaki}, {Dewangan}, {Fabian},
  {Fukazawa}, {Gallo}, {Griffiths}, {Inoue}, {Kunieda}, {Markowitz}, \&
  {Miniutti}}]{reeves07}
{Reeves} J.~N., {Awaki} H., {Dewangan} G.~C., {Fabian} A.~C., {Fukazawa} Y.,
  {Gallo} L., {Griffiths} R., {Inoue} H., {Kunieda} H., {Markowitz} A.,
  {Miniutti} G., 2007, \pasj, 59, 301

\bibitem[{{Reeves} {et~al.}(2005){Reeves}, {Pounds}, {Uttley}, {Kraemer},
  {Mushotzky}, {Yaqoob}, {George}, \& {Turner}}]{reeves05}
{Reeves} J.~N., {Pounds} K., {Uttley} P., {Kraemer} S., {Mushotzky} R.,
  {Yaqoob} T., {George} I.~M., {Turner} T.~J., 2005, \apjl, 633, L81

\bibitem[{{Reynolds} \& {Nowak}(2003)}]{reynolds03}
{Reynolds} C.~S., {Nowak} M.~A., 2003, \physrep, 377, 389

\bibitem[{{Rosenthal}(1979)}]{rosenthal79}
{Rosenthal} R., 1979, {Psy. Bull.}, 86, 638

\bibitem[{{Scargle}(2000)}]{scargle00}
{Scargle} J.~D., 2000, Journal of Scientific Exploration, 14, 91

\bibitem[{{Sterling}(1959)}]{sterling59}
{Sterling} T.~D., 1959, {J. Am. Stat. Ass.}, 54, 30

\bibitem[{{Stern} \& {Simes}(1997)}]{stern97}
{Stern} J.~M., {Simes} R.~J., 1997, {BMJ}, 315, 640

\bibitem[{{Turner} {et~al.}(2004){Turner}, {Kraemer}, \& {Reeves}}]{turner04}
{Turner} T.~J., {Kraemer} S.~B., {Reeves} J.~N., 2004, \apj, 603, 62

\bibitem[{{Turner} {et~al.}(2002){Turner}, {Mushotzky}, {Yaqoob}, {George},
  {Snowden}, {Netzer}, {Kraemer}, {Nandra}, \& {Chelouche}}]{turner02}
{Turner} T.~J., {Mushotzky} R.~F., {Yaqoob} T., {George} I.~M., {Snowden}
  S.~L., {Netzer} H., {Kraemer} S.~B., {Nandra} K., {Chelouche} D., 2002,
  \apjl, 574, L123

\bibitem[{{Yaqoob} {et~al.}(2003){Yaqoob}, {George}, {Kallman}, {Padmanabhan},
  {Weaver}, \& {Turner}}]{yaqoob03}
{Yaqoob} T., {George} I.~M., {Kallman} T.~R., {Padmanabhan} U., {Weaver} K.~A.,
  {Turner} T.~J., 2003, \apj, 596, 85

\bibitem[{{Yaqoob} {et~al.}(1999){Yaqoob}, {George}, {Nandra}, {Turner},
  {Zobair}, \& {Serlemitsos}}]{yaqoob99}
{Yaqoob} T., {George} I.~M., {Nandra} K., {Turner} T.~J., {Zobair} S.,
  {Serlemitsos} P.~J., 1999, \apjl, 525, L9

\bibitem[{{Yaqoob} {et~al.}(1998){Yaqoob}, {George}, {Turner}, {Nandra},
  {Ptak}, \& {Serlemitsos}}]{yaqoob98}
{Yaqoob} T., {George} I.~M., {Turner} T.~J., {Nandra} K., {Ptak} A.,
  {Serlemitsos} P.~J., 1998, \apjl, 505, L87

\bibitem[{{Yaqoob} \& {Serlemitsos}(2005)}]{yaqoob05}
{Yaqoob} T., {Serlemitsos} P., 2005, \apj, 623, 112

\end{thebibliography}

\bsp

\label{lastpage}

\end{document}